\documentclass[lettersize,journal]{IEEEtran}
\usepackage{amsmath,amsfonts}
\usepackage{algorithmic}
\usepackage{algorithm}
\usepackage{array}
\usepackage[caption=false,font=normalsize,labelfont=sf,textfont=sf]{subfig}
\usepackage{textcomp}
\usepackage{stfloats}
\usepackage{url}
\usepackage{verbatim}
\usepackage{graphicx}
\usepackage{cite}
\usepackage{xcolor}
\usepackage{soul}
\usepackage{xcolor}
\definecolor{lightgray}{gray}{0.92}

\newcommand{\code}[2]{%
  \begingroup
  \setlength{\fboxsep}{2pt}%
  \colorbox{lightgray}{\texttt{\detokenize{#2}}}%
  \endgroup
}

\usepackage{listings}
\usepackage[dvipsnames]{xcolor}

\newcommand\YAMLcolonstyle{\color{red}\mdseries}
\newcommand\YAMLkeystyle{\color{black}\bfseries}
\newcommand\YAMLvaluestyle{\color{blue}\mdseries}

\makeatletter

\newcommand\language@yaml{yaml}

\expandafter\expandafter\expandafter\lstdefinelanguage
\expandafter{\language@yaml}
{
  keywords={true,false,null,y,n},
  keywordstyle=\color{darkgray}\bfseries,
  basicstyle=\YAMLkeystyle,                                 % assuming a key comes first
  sensitive=false,
  comment=[l]{\#},
  morecomment=[s]{/*}{*/},
  commentstyle=\color{purple}\ttfamily,
  stringstyle=\YAMLvaluestyle\ttfamily,
  moredelim=[l][\color{orange}]{\&},
  moredelim=[l][\color{magenta}]{*},
  moredelim=**[il][\YAMLcolonstyle{:}\YAMLvaluestyle]{:},   % switch to value style at :
  morestring=[b]',
  morestring=[b]",
  literate =    {---}{{\ProcessThreeDashes}}3
                {>}{{\textcolor{red}\textgreater}}1     
                {|}{{\textcolor{red}\textbar}}1 
                {\ -\ }{{\mdseries\ -\ }}3,
}

\lst@AddToHook{EveryLine}{\ifx\lst@language\language@yaml\YAMLkeystyle\fi}
\makeatother

\newcommand\ProcessThreeDashes{\llap{\color{cyan}\mdseries-{-}-}}

\begin{document}

\title{In QKD, Key Metadata is Key}
% Connecting Elaborated Federal QKD Networks
% CEF-QCI: Coordinating the Execution of Federated Quantum Communication Infrastructures
% noi oferim o arhitectura de orchestrare a conectivitatii de retele elaborate QKD

\author{Alin-Bogdan Popa and Pantelimon George Popescu ~\IEEEmembership{}
        % <-this % stops a space
\thanks{Corresponding author: P.G.Popescu}
\thanks{A.B.Popa and P.G.Popescu are with National University of Science and Technology POLITEHNICA Bucharest (emails: \underline{alin\_bogdan.popa@upb.ro}, \underline{pgpopescu@upb.ro})}% <-this % stops a space
\thanks{Manuscript received XXXXX; revised YYYYY.}}

\markboth{Journal of \LaTeX\ Class Files,~Vol.~14, No.~8, August~2021}%
{Shell \MakeLowercase{\textit{et al.}}: A Sample Article Using IEEEtran.cls for IEEE Journals}

\IEEEpubid{0000--0000/00\$00.00~\copyright~2021 IEEE}

\maketitle

\begin{abstract}
Federated interconnected QKD networks are becoming the norm, and with several standards already in place, interoperability is now technically feasible. Many aspects related to cross-domain key distribution, however, are left unresolved: breach announcement and mitigation, integration with the space segment, extension to a future QKD-as-a-Service commercial model, custom requests referring to key freshness, jurisdictions, security levels, etc. We argue that, to move from syntactic to semantic interoperability, a common language must be developed to ensure transfer of meaning along with the keys. We posit this is possible via QKD key metadata, and we propose a 6-step process to develop it as a standard.
\end{abstract}


\section{Introduction}
\IEEEPARstart{Q}{uantum} computing technology has advanced significantly in recent years. The current (and very recent) record is held by Caltech, with an optical tweezers system capable of trapping 6,100 neutral atom qubits, maintained in coherence for more than 12 seconds, and with single-qubit gate fidelity of more than 99.98\% \cite{manetsch2025tweezer}. Although it has been envisioned until recently that running Shor's algorithm against encryption algorithms used in practice will require at least 400,000 qubits \cite{mckinsey2024quantumadvantage}, theoretical advancements have shown that leveraging high-rate quantum error-correcting codes and efficient design, Shor's algorithm can be executed at relevant scales with as few as 10,000 qubits provided they are reconfigurable enough \cite{cain2026shor}; what is more, theoretical advancements in distributed quantum gates has shown that even systems with much fewer qubits, if supporting distributed quantum computing, can be used to break encryption \cite{tuanuasescu2022distribution}. The so-called "Q-Day" seems now on the horizon.

Quantum key distribution (QKD) has emerged as a main enabler of information-theoretically secure (ITS) communication. Considered more secure than post-quantum cryptography (PQC) and much more scalable than manual, cold-stored one-time pad tapes, QKD is viewed as the prime candidate for future-proof highly-sensitive communication. The European Quantum Communication Infrastructure (EuroQCI) \cite{europeancommission2024euroqci}, the largest QKD initiative in Europe, has had all EU Member States deploying national QKD networks in preparation for the future governmental and international secure communication. The second phase of EuroQCI, currently undergoing development under the CEF Digital (Connecting Europe Facility) funding programme, seeks to interconnect the national QKD networks into an interoperable, pan-European, secure foundation, using terrestrial cross-border QKD links as well as satellite-relayed QKD key exchange for long-distance connections.

One shortcoming of QKD is the limited range of QKD devices (typically, on the order of 100 kilometers). With no practical quantum repeaters and quantum memories as of today, QKD deployments extend individual links into complex networks via the trusted node primitive. Trusted nodes are responsible for relaying keys in order to extend the range of single links, with multiple relaying methods described in the literature and standards, such as ITU-T Y.3803 \cite{itu2023y3803amd1}. Within a single domain (e.g. a national network), keys produced by the QKD devices are typically fetched or streamed via APIs following established standards such as ETSI GS QKD 014 \cite{etsi2019qkd014} and ETSI GS QKD 004 \cite{etsi2020qkd004}. At the Key Management System (KMS) level, various methods can be applied to further increase the key availability and utility, such as by modelling key routing as a constraint optimization problem to increase overall key rates \cite{popa2024optimal}, or by separating series of QKD links into dedicated lanes (virtual links) for better key control via virtualization \cite{popa2024future}.

While mechanisms for key forwarding in an autonomous QKD network have been standardized and demonstrated, interconnecting QKD domains remains an open challenge due to the complexity of the orchestration process \cite{popa2024cef}. The ETSI GS QKD 020 standard \cite{etsi2026qkd020}, published in June 2026, is a good first step: it defines the interface with which different domain KMS systems (e.g. two national networks) can interact horizontally within a boundary trusted node (e.g. on the border of the countries), in order to request, fulfill, fetch, and void QKD-generated keys. At its core, the standard is based on the same key containers already in effect through previous standards like ETSI GS QKD 014: a (set of) base64-encoded binary key(s) associated with a UUID key ID with which it can be recovered at the other end.

\IEEEpubidadjcol

However, beneath the surface, unresolved matters persist. In the following, we will outline several gaps between the key-value/identifier established abstraction and what is needed for practical, scalable deployments.

% The contributions of this work are as follows. First, it derives cross-domain metadata requirements from trusted-node incident response, policy-constrainted key services, and QKD-as-a-Service scenarios. Second, it proposes a signed event graph in which key generation, transformation, transfer, assignment and use are represented as authenticated, auditable operations, while domains retain control over the views exported to other parties. Third, it outlines how metadata requests, views and references can be carried through existing ETSI GS QKD 014 and ETSI GS QKD 020 extension mechanisms without replacing established key delivery standards.

The contributions of this work are as follows. First, it derives metadata requirements from trusted-node incident response, resource-scarce operation, policy-constrained services, and QKDaaS scenarios. Second, it introduces a four-level maturity model for cross-domain QKD interoperability and classifies the information required to progress from syntactic to operational interoperability. Third, it identifies the scope of a common QKD metadata interoperability profile across semantics, binding, transport, trust, enforcement, and extensibility.

\section{The perils of minimal standard support - a short but uncomfortable story}

Consider the following short story. Alice and Bob, living in different QKD domains (Wonderland and Bonderland, respectively), are connected via a series of interconnected domains, as represented in Figure \ref{fig:metadatapolicy_crossdomain}. Within each domain, ETSI GS QKD 014 is used as standard API for key supply, in which both Alice and Bob obtain each key in the form of a Key Container JSON (as per ETSI GS QKD 014), with only two mandatory fields:

\begin{figure*}[t]
\centering
\includegraphics[width=\textwidth]{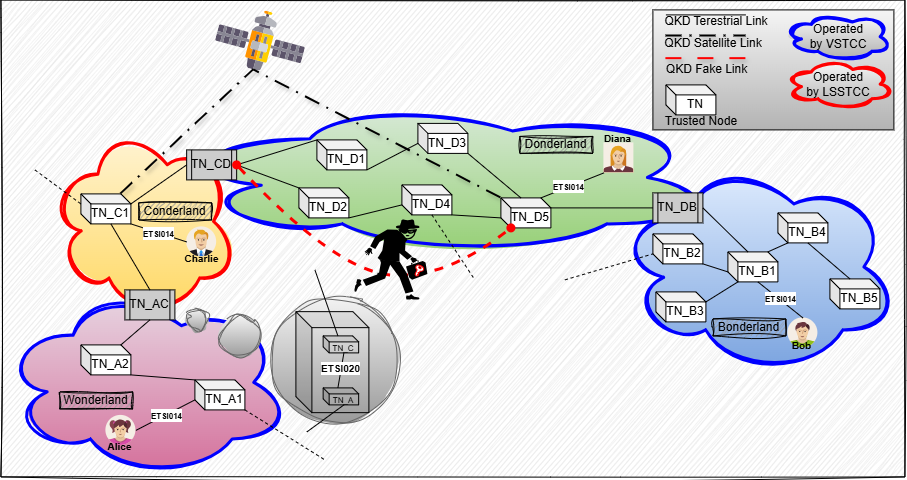}
\caption{An example of a complex cross-domain communication. Four different domains with four users are interconnected: Alice in Wonderland, Bob in Bonderland, Charlie in Conderland, and Diana in Donderland. Conderland and Donderland also benefit from a direct satellite connection. Two QKD service provider companies (represented in Red and Blue outlines) are approved to operate QKD networks in the respective domains. The boundary Trusted Nodes, represented in gray, hold the Key Management systems of both domains, as shown in the zoomed representation next to $TN_{AC}$. The dark silhouette represents a manually carried key via a fake QKD link.}
\label{fig:metadatapolicy_crossdomain}
\end{figure*}

\begin{itemize}
    \item \code{Python}{key}, a string representing the base64-encoded binary value of the final key;
    \item \code{Python}{key_ID}, a UUID uniquely identifying the QKD key between the two ends, and which is safe to transmit over the Internet without assuming a separate secret channel.
\end{itemize}

Domains Wonderland, Conderland, Donderland, Bonderland are interconnected sequentially. Between every two interconnected domains, there is a boundary node, trusted by both domains, which contains the KMS agent of each domain. The two KMS agents interact through an API as defined by ETSI GS QKD 020 standard, which exposes keys in a similar format to ETSI GS QKD 014. While both standards support key container extensions, without an orchestrated agreement between domains, no extensions are used.

Undoubtedly, Alice and Bob are no regular persons; they represent the Government of their respective domains. Their communication is so important that the evil Eve is constantly looking to intercept it. They must communicate often, too: every single day, they have to send important documents to one another. Fortunately, they have the cross-domain QKD network at their disposal: one of them fetches keys, sends the key IDs to the other party so they can re-fetch the same keys on the other side, and they use the keys to encrypt their secret files unconditionally-secure via One-Time Pad. And all is good.

\subsection{A regrettable mistake}

One day, David, the President of Donderland, makes a disastrous public announcement: last Wednesday, his node $TN_{D2}$ was breached! Eve managed to silently take control over the node between 2pm and 4pm (in the local timezone of Donderland residents, of course) before the automated key rotation system activated and kicked her away. 

For Alice and Bob, this is not something they can casually dismiss. They decide to get together and take a look at the map in Figure \ref{fig:metadatapolicy_crossdomain}. If the keys that they used in their communication took the path $TN_{CD} \rightarrow TN_{D2} \rightarrow TN_{D4} \rightarrow TN_{D5}$, then their communication must have been compromised. If instead the keys took the route $TN_{CD} \rightarrow TN_{D1} \rightarrow TN_{D3} \rightarrow TN_{D5}$, then their communication is still safe. They pose this question to David, but Donderland engineers do not log this information, and so without any sure information, they must assume Eve could have decrypted their communication. Of course, a satellite connection was available between nodes $TN_{C1}$ and $TN_{D5}$ directly, but again, nobody is sure whether satellite-generated keys were being used at the time.

They look in their own logs. Alice, who is perhaps more diligent in keeping paper records (she has logs of which applications were active at which timestamps), breathes a sigh of relief: last Wednesday, between 2pm and 4pm (in the timezone of Donderland), the only secure application that was active between her and Bob was a weather report software. That's good - she thinks - all Eve could have decrypted was some temperature values and weather forecasts. Her relief is not long lasting, though, as she is soon informed by her personal technical assistant that in QKD networks, keys are not actually used as soon as they are generated. Contrary to popular belief - he says - keys are generated, then stored in temporary link-buffers, then moved to consumer pair buffers, then collected by the KMS for forwarding, then sent to the requesting parties, and then used - and this process may take time. Hence, there is a significant chance that a key generated in one link ending at the affected $TN_{D2}$, could have been used for Alice's and Bob's communication taking place at 7pm, when Bob sent Alice his nuclear launch codes, or at 8pm, when Alice sent Bob her yearly national security plans.

With no way to assess the severity and extent of the damage, they are left no choice but to revise their national security plans, change the nuclear launch codes, and overhaul systems within entire branches of their governments whose communication was potentially compromised in the few days following the attack. Both Alice and Bob are left with a bitter taste regarding QKD, and they vow to never use it again. If only there was a way to track a key from one end to the other!

\subsection{One man's key is another man's national security}

Hearing about the recent security incident in Donderland, Charlie (from Conderland, of course) and Diana ponder about the current status of cross-domain interconnection. Luckily, as Heads of the national security services in their respective domains, they were fortunate enough to have been able to convince their governments to invest a significant amount of funds into developing Optical Ground Stations (OGSs). The International QKD Satellite ensures a direct QKD connection with any OGS over large distances, bypassing terrestrial, potentially vulnerable routes. So when an important national security communication gets scheduled to take place the next morning, they are not alarmed: the file would be transferred directly from node $TN_{D5}$ to node $TN_{C1}$, using keys that were exchanged via satellite directly. 

Just to be sure, Diana decides to call the QKD engineers. Imagine her surprise when she learns there are no satellite keys left in node $TN_{D5}$! How could this happen? Well - her engineer explains - keys requested through node $TN_{D5}$ (perhaps by other Donderland officials, but also perhaps by Alice and Bob) during a high-traffic burst, consumed the full key buffer of node $TN_{D5}$. With no keys left and no standing policy or other mechanism to tell the difference between terrestrial and satellite links (as established, the satellite connection is treated as just another QKD link, no different than the others), the satellite keys go into the same key pool as the rest, and they have also been consumed. Can't we just get another key? Diana asks. Unfortunately - the engineer says - that is impossible: the satellite is a scarce resource, its useful passes are limited to only few per day, the weather conditions are not good, and the key request must be scheduled 5 days in advance anyway. There is simply not enough time. Then, Diana asks, if the satellite key is so difficult to obtain and its infrastructure is so expensive, why is it treated the same as a terrestrially-exchanged key? The engineer does not have an answer.

Diana ends her phone call and stares at her still uneaten food. On her table there are two shakers, identical in every aspect except one is clearly labeled "SALT", while the other is labeled "PEPPER". She has never put pepper accidentally in her food when she meant to put salt. She has always checked the labels. An idea starts to form as a quiet spark in her mind.

\subsection{The wealth of networks}

VSTCC, a Very Successful TeleCommunications Company with a well-chosen name, has offices in all the domains from our story. As one of the main players on a global scale, they provide high-quality internet connections, and having many cybersecurity and defense certifications, they do not shy away from contracts with clients interested in high-end security, such as governmental agencies, military facilities, banks, and many more.

Following recent developments, VSTCC becomes very interested in QKD networks. Their clients are always asking for increased security of communications; with the advancements in quantum computing, it is time to integrate quantum key distribution. The world is moving to large-scale adoption. However, for VSTCC it would be quite infeasible to built a private, parallel QKD network from scratch; QKD devices, nodes, and their security, all are still very expensive. They have a different plan: they want to sign contracts with the governments of all domains, allowing them to purchase a dynamic key rate quota across the national QKD backbones of the domains. They would then sell/deliver the keys to governmental agencies or private companies connected to the nodes, as part of a QKD as a Service (QKDaaS) approach. The plan is flawless: the government receives additional money for generated keys that would otherwise expire unused, QKD adoption increases, VSTCC has leftover money with which they can extend the national QKD networks, and general communication can be done in a more secure manner. Everyone signs the contract without a second thought.

Soon enough, VSTCC's development hits unexpected blockers. A bank headquartered near Wonderland node $TN_{A2}$ has a subsidiary near $TN_{B3}$ in Bonderland. As a client of VSTCC, they ask for a dynamic retainer of keys between the two locations. VSTCC engineers try to find a solution: they need all nodes across the four domains to reserve a variable amount of keys whose ownership must be attached to this tenant; moreover, they need exact information on what was reserved, for how long, and across which node, for billing purposes. The deployed devices and KMS solutions offer no technical mechanism for programmatic reservations, nor a standardized export of machine-readable information regarding key ownership. Without a choice, VSTCC strikes a deal with each domain individually: the domain engineers are to implement such mechanisms specifically for VSTCC, following VSTCC's proprietary standard. There is a problem, though, because in Conderland, LSSTCC, a Local Somewhat Successful TeleCommunications Company, has already imposed their own proprietary key ownership standards across Conderland. Since the companies are unwilling to change their standards for each other (perhaps they have different billing mechanisms or submit to different regulations), the conderlandian government must now support multiple standards in their network, in commercial interest. With every addition, the network grows more fragmented.

A popular joke goes like: a quality assurance engineer walks into a bar and orders 1 beer. Then, he orders 2 beers. Then, 1.5 beers, 0 beers, -1 beer, 999 million beers. The first real customer walks into the bar and asks where the bathroom is; the bar explodes. The experience of VSTCC is quite similar in that regard. They expected a smooth process in which keys are generated end-to-end and clients consume the keys. Instead:
\begin{itemize}
    \item One client requests keys between $TN_{A1}$ and $TN_{B1}$ which specifically avoid in their routing any node within the entirety of the Donderland. VSTCC would need custom bilateral agreements with all domains and custom engineering solutions to be able to relay such requests.
    \item A bank requested keys that were generated only via devices manufactured by a vendor from an approved list, AND only through nodes whose devices have a software firmware that has been updated less than 2 months ago, AND only provided that forwarding from one device to the next is done by encrypting keys specifically with OTP, not with AES. VSTCC would need, again, custom bilateral agreements and audits with each domain; however, the way keys are encrypted is done opaquely at a device/KMS configuration level and may not be updated on the fly; VSTCC begins discussion with all QKD manufacturers in a futile attempt to reach an agreement.
    \item A corporate client aggregates keys in a vault and demands a Service-Level Agreement in which keys are contractually fetched at the moments where traffic is lower and keys are less expensive. To fulfil this, VSTCC would need exact information on key usage demand across all domains, something that is technically difficult and perhaps politically infeasible due to impact on national security of respective domains.
    \item A national security facility requires keys which were generated end to end (including all their component keys) no longer than 10 seconds before they are received and used. VSTCC realizes they can only control the time of request, but they have no way to check or ensure the received key's freshness.
\end{itemize}

With so many issues at play, VSTCC takes a step back and announces that the launch of the new global QKDaaS infrastructure is postponed indefinitely, as the technology is simply not mature enough. 

\subsection{Is this the real life?}

Seeing the recent security attack at node $TN_{D2}$, Donderland officials decide to hire a team of experts to implement the security best practices at each of their nodes. The experts, highly trained in the ancient arts of cybersecurity, quickly come up with a long (and expensive) list of changes: default admin credentials are to be removed, devices are to be reconfigured, access control is to be overhauled, and so on. A minor discovery during the audit also shows that node $TN_{D3}$ has never really worked properly, but its warranty has long expired and the reservice cost is outside Donderland's budget for the year. Not a problem - so they think - they will reconfigure the route $TN_{CD} \rightarrow TN_{D2} \rightarrow TN_{D4} \rightarrow TN_{D5}$, at least.

After a lengthy hard work on reconfiguring the nodes, the job is finally done and the network is restarted. But as fate would have it, no keys are produced along the reconfigured segment. What could have happened? Perhaps a misconfigured initial key, maybe a faulty device; everything is on the table. The experts inform the Donderland government that debugging the network and reconfiguration may take another month.

Usually, this would not be a problem. In this case, however, Donderland has signed bilateral agreements with the domains of Conderland and Bonderland to allow up to 20\% of their internal QKD key material for cross-domain, international use-cases. As part of the same bilateral agreements, Donderland has also committed to fixing any discontinued service issue in a maximum of 3 days. Charlie and Bob have already made a failed attempt to exchange a QKD key, they raised the issue to their respective national agencies, and Donderland expects to receive an official demand for service soon.

Thinking quickly, David takes a difficult (but not entirely unforseable) decision. Donderland is to install an alternative system in trusted nodes $TN_{CD}$ and $TN_{D5}$. The system would generate random strings as key material (using a quick, classical pseudo-random number generator instead of the quantum counterpart) in $TN_{CD}$, send it via an unsecured channel to $TN_{D5}$ (for example, via E-mail, or via an unprotected API), and then expose it to the rest of the QKD network using the standard interfaces. To an external node, the key is just a string of random bits: it has no clear identity, no authenticated route/forwarding history, no guarantee that it ever came from a QKD device in the first place. Nobody would know, and life would go on.

Diana, who, like most of the quantum community, is highly involved in the technical matters of her government and is made aware of this development, takes an unfocused glance at her recently acquired QKD keys with Alice. If her government is willing to install what is essentially a fake QKD link just to play pretend there is a connection, perhaps others are doing the same? She starts to wonder whether she can really trust any QKD key that came from anyone except from her own device in her own lab; hard science and security proofs are worth nothing to her if the implementation is all but a farce. Of course - she thinks - this could have easily been avoided with some mechanism that could guarantee authenticity. An expensive bottle of perfume, a naturally-sourced diamond engagement ring, a luxury guitar and an original Monet painting, all have in common the fact they usually come with a certificate of authenticity, and somewhere out there is an authority capable of validating it. She looks again at her QKD keys, this time slightly annoyed. If only!

\section{Label your keys!}

\begin{figure}[ht]
\centering
\includegraphics[width=0.35\textwidth]{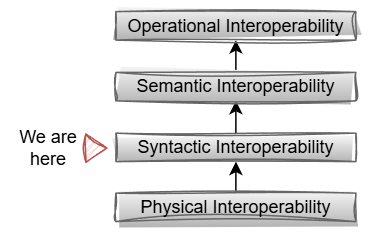}
\caption{The four levels of QKD cross-domain interoperability}
\label{fig:metadatapolicy_interoperability}
\end{figure}

For a federated, cross-domain QKD deployment, we define a separation of interoperability into several levels as described in Figure \ref{fig:metadatapolicy_interoperability}. The first level and the easiest to implement practically as demonstrated throughout various deployments, Physical Interoperability, requires QKD links to exist in a cross-domain fashion. The second level (where we argue the current state of affairs is) is Syntactic Interoperability: at this level, standard APIs exist (such as ETSI GS QKD 014, ETSI GS QKD 020) which allow nodes and key management systems to perform the minimal actions in order to successfully exchange keys intra- and inter-domain. At the third level, Semantic Interoperability, different domains' KMSs have a common language and meaning attached to data exchanges; that is, different domains agree on what the exchanged information means. At the fourth and final level, Operational Interoperability, policies and entitlements are fully supported end to end, various SLAs can be guaranteed and enforced, incidents can be mitigated and controlled, and audits can be operationalized on a global scale.

To achieve the next stage in cross-domain QKD interoperability, nodes and domains must be able to exchange semantic information besides the raw keys. In the remainder of this paper, we will collectively call all such semantic information "metadata", whereas when such a piece of information can be meaningfully attached to a particular key, we will call it, more specifically, a "label".

Metadata can provide significant benefits in a complex QKD network, many of which we can derive from the previous story scenarios:
\begin{itemize}
    \item In assessing the security level of keys or damage controlling a security incident, metadata can provide answers to questions such as: Which application was using which key and at what time? In the process of constructing a relayed key, what path did the key take, which nodes and links were involved, and at what time was it relayed/forwarded by each node? In the forwarding process, which mechanism was used, and what type of encryption? Which manufacturer, model, firmware, configuration, and certification profile of QKD devices were used in this forwarding, to what security level are each of them certified, and, perhaps more importantly, who performed this certification?
    \item Regarding resource scarcity: How valuable or scarce is a particular key? Did it use a contested link that is rarely available (for example, a satellite connection)? And, if yes, under what parameters was it generated? How many such keys are still available? Is the currently requesting application important or sensitive enough to use such a key?
    \item In a commercial QKDaaS system, one can define a complex ownership model (such as: infrastructure owner, service beneficiary, operational custodian, authorized consumer, billing party, entity permitted to delegate entitlement, etc.). Metadata could answer questions such as: which entity fulfills each of these roles? If an SLA includes clauses such as key freshness, forwarding method, security or assurance profile of nodes, Quality of Service guarantees, how to assess each relevant metric for a particular key, in order to not breach SLA and to accurately price key material? Moreover, how to allow dynamic client requests for specific metrics, how to show evidence as to whether they were fulfilled, and how to preserve this information in cross-domain or hierarchical subletting key transit?
\end{itemize}

The concept of metadata and per-key labels is anything but new. The ITU-T Y.3803 standard \cite{itu2023y3803amd1} defines basic information for metadata that could be included in the key container, as well as in the Key Management Agent (KMA) container and the Key Supply Agent (KSA) container, where in this terminology the KMA is the software entity responsible for key forwarding and relay between nodes and for managing the key lifecycle, and the KSA is the entity responsible for providing the key to the applications via a key supply interface (for example, ETSI GS QKD 014). The standard suggests metadata elements, most of which marked as optional, such as: module IDs of the QKD devices, generation timestamps, hash values of the secret key data, relay timestamp and relay encryption method, details about the source and destination requesting applications, and even a hierarchical metadata of the underlying key before relay (which is proposed as mandatory). ITU-T FG QIT4N D2.3 Part 2 \cite{itut2021fgqit4nd23part2} further defines the data structures and protocol messages for key relay, authentication, events, and more, including these metadata elements. A practical deployment covering many of the enumerated elements has been performed in the Tokyo QKD Network testbed by Tajima et al. \cite{tajima2017quantum}. The ETSI GS QKD 015 standard \cite{etsi2022qkd015} proposes interfaces to be exposed by the Software Defined Network (SDN) controllers, in terms of network, routes, and applications. Other standardization initiatives recommend metadata to be included especially for cross-domain interoperability, without providing explicit descriptions of interfaces. For example, the recommendation ITU-T Y.3813 \cite{itu2024y3813}, recommends the KMS layer to exchange metadata such as key ID, QKD module ID and key generation rate, with optional information as to which KM the key is transferred, timestamp, cryptographic application to which the key is supplied, shared key number of a KM link, key consumption rate, KM link status (Req\_KM 3); it further recommends the QKD Network control layers to share information such as routing control information, session control information, authentication control information, and quality of service policy control information (Req\_C 1), charging policy control (Req\_C 2), and routing control information such as QKD node addresses, key manager IDs, key consumption rate, and residual number of keys (Req\_C 4).

QKDaaS poses further challenges due to the added complexity of SLAs. In Cheng et al. \cite{cheng2011qos}, a QoS-supported key manager is proposed, in which three service classes are proposed: key-guaranteed service, key-prioritized service, and key-best-effort service, differentiated by distribution time (defined as the total processing time required for a key to travel from its source to its destination), using additional key-data to differentiate the service classes. Similarly, Horoschenkoff et al. \cite{horoschenkoff2025demoquandt} presented a QKD network architecture, deployed between Berlin and Bonn in Germany, with a clear separation between user KMS, access KMS, and carrier KMS - an important distinction towards QKDaaS - by enriching requests at each level with data such as the number and size of keys, the participating countries, and the contractual context that enables verification (e.g. payment, service level). A more comprehensive survey of KMS solutions and proposals can be found in \cite{dervisevic2025quantum}. Preserving metadata during key transit, particularly cross-domain, is also not an easy task. Sometimes keys are not merely forwarded, but may be split and recombined in various ways, in which case metadata of all underlying keys should be represented in the final key. James et al. \cite{james2023key} provide a useful KMS API proposal covering operations such as new\_key\_batch, forward\_keys, split\_key, merge\_keys, delete\_keys, make\_keys\_internal, make\_keys\_external, operations which can be extended to compose metadata into a key lineage graph.

The addition of key metadata has an obvious, but significant drawback: it increases the attack surface and potentially exposes information about the network that should remain secret. Although some metadata elements may remain public without a significant security impact (this should be assessed as future work), many metadata elements may need to be transmitted in encrypted form. ITU-T Y.3803 \cite{itu2023y3803amd1} proposes key metadata to be encrypted via XOR (One-Time Pad) to preserve the ITS property, although we expect this approach to significantly increase the QKD key rate usage for a non-direct key relay (this analysis has not been performed yet, to our knowledge). ITU-T X.1712 \cite{itu2021x1712} further details security requirements on key metadata, covering encryption, integrity, and authentication. Brauer et al. \cite{brauer2024linking} added metadata information, including a key validity period and the name of the sending node, signed using PQC algorithms at the KMS level, in a simulated long-distance linking of three European QKD testbeds, in Berlin, Madrid, and Poznan. Towards limiting information exposed about networks, there is also significant progress: Krenn et al. \cite{krenn2026topology} proposed a zero-knowledge proof architecture in which nodes can prove compliance with agreed-upon policies (e.g. a minimum node certification level or a QKD device provenance) along the forwarding route, without exposing the exact network topology or nodes involved.

It is of note that common key supply and transition standards such as ETSI GS QKD 014 and ETSI GS QKD 020, mechanisms for mandatory and optional extensions are supported. Such mechanisms could be used to request, relay, and reply to, metadata information without replacing established key delivery standards.

With the above precedents in deployments, theoretical advancements, and partial support in standards, we argue that the missing piece towards Semantic Interoperability is now a common QKD metadata profile. In the following section, we derive the requirements for such a profile and propose a methodology for establishing it.

\section{The way forward}

\subsection{Metadata classes}

We start by distinguishing five classes of metadata:
\begin{enumerate}
    \item Key-associated context - Metadata which describes the origin and handling of key material. Some examples include: generation source, timestamp, key composition graph, transformations, encryption method during forwarding, paths taken, etc.
    \item Service/request metadata - Metadata which describes the specific requirements of the requesting parties that the network should provide or satisfy. Some examples include: jurisdiction constraints, assurance or certification classes, priority, key freshness, required forwarding or encryption methods, request for specific key-associated context information, etc.
    \item Entitlement and accounting metadata - Metadata which describes key rights and economic value. Some examples include: beneficiary, custodian, requesting application, quota, rights delegation, SLA class, billing information, etc.
    \item Resource and inventory metadata - Metadata which describes key inventory. Examples include: key scarcity, expiry, reserved/held/committed/consumable quantities, replenishment rate, predicted usage, etc.
    \item Evidence and assurance metadata - Metadata which provides factual evidence of claims. Some examples include: compliance proofs, node/domain signatures or certificates, audit references, etc.
\end{enumerate}

This separation is necessary because different types of metadata may require different handling: key-associated context can be handled at the level of nodes and domain SDN controllers; service/request metadata is attached only to request, not to specific keys; entitlement and accounting metadata lives at a higher abstractization level, and, depending on the intended usage of the network (e.g. if it is never meant to be used commercially or in a QKDaaS model), may be entirely optional; the resource and inventory metadata, already partially supported in established standards such as the Status endpoint defined in ETSI GS QKD 014, requires critical extensions to better serve resources in scarcity-driven environments; and the evidence and assurance metadata is the glue that holds the other types together and provides the means of verification and metadata integrity.

\subsection{A standardization roadmap}

\begin{figure}[ht]
\centering
\includegraphics[width=0.35\textwidth]{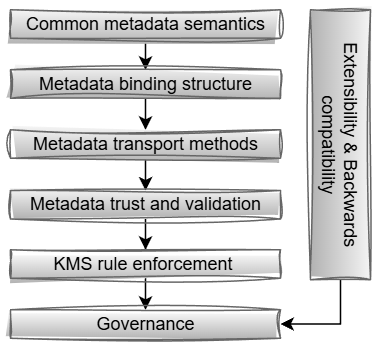}
\caption{A roadmap towards a common QKD metadata profile}
\label{fig:metadatapolicy_roadmap}
\end{figure}

We present in Figure \ref{fig:metadatapolicy_roadmap} a 6-step roadmap towards a common QKD metadata profile. Below we describe the steps one by one.

A common QKD metadata profile must first define the common semantics, such that all participants agree on the meaning of concepts involved: origin, generation time, supply time, key components, custody, beneficiary, jurisdiction, service class, delegation, and more. Examples of failure criteria include: one node attaches to a key as "timestamp" the local time when the key was generated, whereas another node attaches as "timestamp" the local time when the generation event was recorded in a local database; or, two nodes apply the same OTP-based underlying protocol in the encryption of a forwarded key, but one node calls this process "otp" and the other calls it "xor".

The second layer of a QKD metadata profile is the binding structure. The profile must define how metadata is linked unambiguously with other entities such as: a raw key or key id, a service request, a multi-step transaction/session, a node or domain, an audit inquiry, etc. Moreover, where applicable, the profile should define the cross-reference of the binding structure. For example, keys may be split, joint, forwarded using further keys, combined (in a multipath or k-of-n secret sharing-style), derived via Key Derivation Functions, and more. The binding structure should specify how this graph-like set of operations is represented, and how underlying metadata is referenced.

The third layer of the metadata profile is the metadata transport. While existing ETSI extension mechanisms can be metadata carriers, the model should not depend exclusively on one API. Metadata may be carried inline (as part of the key request/response containers; it may be carried only as a reference that can be later retrieved through an API; it may be transported through orchestrated, domain-level interface administered centrally and potentially recorded in global registries; or it may be supplied later, upon request, through a governmental channel. The metadata profile should strive for transport independence.

The fourth layer of the metadata profile is the trust layer. The profile must establish which entities hold authoritative rights to metadata: potentially, the nodes, the applications, the KMSs, the vendors, the domain central controllers, etc. The profile must further define who may assert which claim; how claims and metadata are authenticated; whether both sides of a transaction must acknowledge a claim, and if yes, whether this process must be interactive or non-interactive; how vendors, nodes, domains demostrate compliance; how a recipient evaluates a claim. Of particular importance is selective disclosure: a domain may wish to expose data ranging from complete information, to restricted summaries, to encrypted details, or even to privacy-preserving zero-knowledge proofs. The metadata profile should allow the generation and verification of such claims without requiring national domains to reveal the full internal topology, nor the full chain of key operations.

At the fifth layer, the QKD metadata profile provides guidelines for rule enforcement at KMS level. If the forwarding paths, the encryption methods, or the key derivation mechanisms performed at a node level are dependent on metadata associated to underlying keys or to the end-user's request, then the nodes, KMSs, boundary agents, and domains may need to be aware of the actions to take when the keys and requests reach them. The metadata profile should provide a general, composable and extensible rule methodology that device and KMS vendors can implement.

Within each of these first 5 layers, the metadata standard must permit versioned fields and profiles, optional claims and extensions, mandatory compliance, handling guidance for cases where a domain or node does not understand a specific field or request (including, perhaps, commands such as: full forward, drop metadata, abort request, etc.), and algorithm agility where applicable.

Finally, the metadata profile must specify its governance rules. It should clarify who maintains namespaces and event/field registries, what the mandatory baseline is and which elements are optional, how are improvements added (whether centralized or decentralized), what the approval process is for new metadata profiles or old version deprecation, what interoperability tests are required, and who resolves disputes when different actors disagree over claims or historical records.

\subsection{Why the time is now}

The common QKD metadata profile must be defined before networks are built and interconnected, for several reasons.

First, the naive approach based around custom bilateral agreements does not scale. In a network with $N$ independent domains, the number of bilateral agreements grows as $N^2$. Moreover, any end-to-end service is only as interoperable as its least compatible domain along the graph path. The relevant actors, however, are not limited to domains: QKD vendors, satellite controllers, telecommunication companies, all may impact (or block) the result. While existing extension mechanisms enable proprietary metadata easily, without a common framework, actors will not converge spontaneously to a standardized format; in this regard, proprietary extensions may increase lock-in and fragment the networks further rather than solve federation.

Second, and more important, historical information can never be reconstructed. Even if metadata APIs are retrofitted to existing network, relevant information that was never recorded and stored cannot be recreated after a key has been consumed, expired, or destroyed. With every passing day that QKD networks are in operation, the amount of potential data that could have been stored and made available for audit, technical integrations, or incident mitigation, increases. We consider this metadata debt.

Since the metadata profile may influence many aspects from procurement to architecture (for example: node storage, event logging, global identifiers methodology, certificate and attestation infrastructure, inter-node and inter-domain interfaces, OGS and satellite session records, etc.), we argue that the common QKD metadata profile is time-critical before QKD networks, especially in federated domains, can achieve desired levels of interoperability.

\section{Conclusion}

In this paper we describe several scenarios that pose as significant blockers towards scalable cross-domain QKD network deployments, in particular related to security issue mitigation, QKD as a Service, and custom key or node requirements. We show how QKD key metadata, with useful precedents in standards and literature, can be a solution to these problems, and we propose a high-level operational framework as a starting point towards the natural next step of semantic interoperability in federated QKD networks.

\section*{Acknowledgement}

% \section{References Section}
 

\bibliographystyle{IEEEtran}
\bibliography{bib}

@article{manetsch2025tweezer,
  title={A tweezer array with 6,100 highly coherent atomic qubits},
  author={Manetsch, Hannah J and Nomura, Gyohei and Bataille, Elie and Lv, Xudong and Leung, Kon H and Endres, Manuel},
  journal={Nature},
  volume={647},
  number={8088},
  pages={60--67},
  year={2025},
  publisher={Nature Publishing Group UK London}
}

@article{tuanuasescu2022distribution,
  title={Distribution of controlled unitary quantum gates towards factoring large numbers on today’s small-register devices},
  author={T{\u{a}}n{\u{a}}sescu, Andrei and Constantinescu, David and Popescu, Pantelimon George},
  journal={Scientific Reports},
  volume={12},
  number={1},
  pages={21310},
  year={2022},
  publisher={Nature Publishing Group UK London}
}

@article{cain2026shor,
  title={Shor's algorithm is possible with as few as 10,000 reconfigurable atomic qubits},
  author={Cain, Madelyn and Xu, Qian and King, Robbie and Picard, Lewis RB and Levine, Harry and Endres, Manuel and Preskill, John and Huang, Hsin-Yuan and Bluvstein, Dolev},
  journal={arXiv preprint arXiv:2603.28627},
  year={2026}
}

@misc{mckinsey2024quantumadvantage,
  author       = {{McKinsey \& Company}},
  title        = {Steady Progress in Approaching the Quantum Advantage},
  howpublished = {\url{https://www.mckinsey.com/capabilities/mckinsey-digital/our-insights/steady-progress-in-approaching-the-quantum-advantage}},
  year         = {2024},
  note         = {Accessed: June 10, 2024}
}

@misc{europeancommission2024euroqci,
  author       = {{European Commission}},
  title        = {The European Quantum Communication Infrastructure ({EuroQCI}) Initiative},
  howpublished = {\url{https://digital-strategy.ec.europa.eu/en/policies/european-quantum-communication-infrastructure-euroqci}},
  year         = {2024},
  note         = {Accessed: June 10, 2024}
}

@techreport{itu2023y3803amd1, title = {Quantum Key Distribution Networks---Key Management: Amendment 1}, author = {{International Telecommunication Union}}, institution = {International Telecommunication Union}, type = {{ITU-T} Recommendation}, number = {Y.3803 (2020) Amd. 1}, month = nov, year = {2023}, url = {https://handle.itu.int/11.1002/1000/15715} }

@techreport{etsi2019qkd014, author = {{European Telecommunications Standards Institute}}, title = {Quantum Key Distribution ({QKD}); Protocol and Data Format of {REST}-Based Key Delivery {API}}, institution = {European Telecommunications Standards Institute}, type = {{ETSI} Group Specification}, number = {ETSI GS QKD 014 V1.1.1}, month = feb, year = {2019}, url = {https://www.etsi.org/deliver/etsi_gs/QKD/001_099/014/01.01.01_60/gs_qkd014v010101p.pdf} }

@techreport{etsi2020qkd004, author = {{European Telecommunications Standards Institute}}, title = {Quantum Key Distribution ({QKD}); Application Interface}, institution = {European Telecommunications Standards Institute}, type = {{ETSI} Group Specification}, number = {ETSI GS QKD 004 V2.1.1}, month = aug, year = {2020}, url = {https://www.etsi.org/deliver/etsi_gs/QKD/001_099/004/02.01.01_60/gs_qkd004v020101p.pdf} }

@article{popa2024optimal,
  title={Optimal key forwarding strategy in QKD behaviours},
  author={Popa, Alin-Bogdan and Popescu, Pantelimon George},
  journal={Scientific Reports},
  volume={14},
  number={1},
  pages={13977},
  year={2024},
  publisher={Nature Publishing Group UK London}
}

@article{popa2024future,
  title={The future of QKD networks},
  author={Popa, Alin-Bogdan and Popescu, Pantelimon George},
  journal={arXiv preprint arXiv:2407.00877},
  year={2024}
}

@article{popa2024cef,
  title={CEF: Connecting Elaborate Federal QKD Networks},
  author={Popa, Alin-Bogdan and Popescu, Pantelimon},
  journal={arXiv preprint arXiv:2409.12027},
  year={2024}
}

@techreport{etsi2026qkd020, author = {{European Telecommunications Standards Institute}}, title = {Quantum Key Distribution ({QKD}); Protocol and Data Format of {REST}-Based Interoperable Key Management System {API}}, institution = {European Telecommunications Standards Institute}, type = {{ETSI} Group Specification}, number = {ETSI GS QKD 020 V1.1.1}, month = jun, year = {2026}, url = {https://www.etsi.org/deliver/etsi_gs/QKD/001_099/020/01.01.01_60/gs_QKD020v010101p.pdf} }

@article{tajima2017quantum,
  title={Quantum key distribution network for multiple applications},
  author={Tajima, Akio and Kondoh, Takashi and Ochi, Takao and Fujiwara, Mikio and Yoshino, Ki and Iizuka, Hiromi and Sakamoto, Toshio and Tomita, Akihisa and Shimamura, E and Asami, Shione and others},
  journal={Quantum Science and Technology},
  volume={2},
  number={3},
  pages={034003},
  year={2017},
  publisher={IOP Publishing}
}

@inproceedings{james2023key,
  title={Key management systems for large-scale quantum key distribution networks},
  author={James, Paul and Laschet, Stephan and Ramacher, Sebastian and Torresetti, Luca},
  booktitle={Proceedings of the 18th International Conference on Availability, Reliability and Security},
  pages={1--9},
  year={2023}
}

@article{dervisevic2025quantum,
  title={Quantum key distribution networks-key management: A survey},
  author={Dervisevic, Emir and Tankovic, Amina and Fazel, Ehsan and Kompella, Ramana and Fazio, Peppino and Voznak, Miroslav and Mehic, Miralem},
  journal={ACM Computing Surveys},
  volume={57},
  number={10},
  pages={1--36},
  year={2025},
  publisher={ACM New York, NY}
}

@article{horoschenkoff2025demoquandt,
  title={DemoQuanDT: a carrier-grade QKD network},
  author={Horoschenkoff, Peter and Henrich, J and B{\"o}hn, R and Khan, I and R{\"o}diger, Jasper and Gunkel, Matthias and Bauch, M and Benda, J and Bl{\"a}cker, P and Eichhammer, E and others},
  journal={Journal of Optical Communications and Networking},
  volume={17},
  number={9},
  pages={743--756},
  year={2025},
  publisher={Optica Publishing Group}
}

@article{brauer2024linking,
  title={Linking QKD testbeds across Europe},
  author={Brauer, Max and Vicente, Rafael J and Buruaga, Jaime S and M{\'e}ndez, Rub{\'e}n B and Braun, Ralf-Peter and Geitz, Marc and Rydlichkowski, Piotr and Brunner, Hans H and Fung, Fred and Peev, Momtchil and others},
  journal={Entropy},
  volume={26},
  number={2},
  pages={123},
  year={2024},
  publisher={MDPI}
}

@article{krenn2026topology,
  title={Topology-hiding path validation for large-scale quantum key distribution networks},
  author={Krenn, Stephan and Mir, Omid and Lor{\"u}nser, Thomas and Ramacher, Sebastian and Wohner, Florian},
  journal={arXiv preprint arXiv:2604.01831},
  year={2026}
}

@techreport{itut2021fgqit4nd23part2,
  author      = {{ITU-T Focus Group on Quantum Information Technology for Networks}},
  title       = {Quantum Key Distribution Network Protocols: Key Management Layer, {QKDN} Control Layer and {QKDN} Management Layer},
  institution = {International Telecommunication Union},
  type        = {{ITU-T} Technical Report},
  number      = {FG QIT4N D2.3, Part 2},
  month       = nov,
  year        = {2021},
  url         = {https://www.itu.int/dms_pub/itu-t/opb/fg/T-FG-QIT4N-2021-D2.3.2-PDF-E.pdf}
}

@techreport{etsi2022qkd015,
  author      = {{European Telecommunications Standards Institute}},
  title       = {Quantum Key Distribution ({QKD}); Control Interface for Software Defined Networks},
  institution = {European Telecommunications Standards Institute},
  type        = {{ETSI} Group Specification},
  number      = {ETSI GS QKD 015 V2.1.1},
  month       = apr,
  year        = {2022},
  url         = {https://www.etsi.org/deliver/etsi_gs/QKD/001_099/015/02.01.01_60/gs_QKD015v020101p.pdf}
}

@techreport{itu2024y3813,
  author      = {{International Telecommunication Union}},
  title       = {Quantum Key Distribution Network Interworking---Functional Requirements},
  institution = {International Telecommunication Union},
  type        = {{ITU-T} Recommendation},
  number      = {Y.3813},
  month       = sep,
  year        = {2024},
  url         = {https://www.itu.int/rec/T-REC-Y.3813-202409-I/en}
}

@inproceedings{cheng2011qos,
  title={A QoS-supported scheme for quantum key distribution},
  author={Cheng, Xianzhu and Sun, Yongmei and Ji, Yuefeng},
  booktitle={2011 International Conference on Advanced Intelligence and Awareness Internet (AIAI 2011)},
  pages={220--224},
  year={2011},
  organization={IET}
}

@techreport{itu2021x1712,
  author      = {{International Telecommunication Union}},
  title       = {Security Requirements and Measures for Quantum Key Distribution Networks---Key Management},
  institution = {International Telecommunication Union},
  type        = {{ITU-T} Recommendation},
  number      = {X.1712},
  month       = oct,
  year        = {2021},
  url         = {https://handle.itu.int/11.1002/1000/14805}
}

\section*{Annex - How to label a key}

% \section{How to label a key}

We start by laying out explicit requirements for a QKD key metadata system. It should support, at minimum, the following:
\begin{enumerate}
    \item Immutable cryptographic key provenance
    \item Split/merge/relay/derivation composition
    \item Lifecycle and inventory traceability
    \item Jurisdiction and trust-policy evaluation
    \item Verifiable, attributable claims rather than self-asserted
    \item Selective disclosure of sensitive topology
    \item Incident impact analysis
    \item Rights, custody and delegation representation
    \item Transport and vendor independence
    \item Bounded metadata growth
    \item Versioning and algorithm agility
\end{enumerate}

\subsection{Modelling a metadata-aware Key Manager}

\begin{figure}[ht]
\centering
\includegraphics[width=0.5\textwidth]{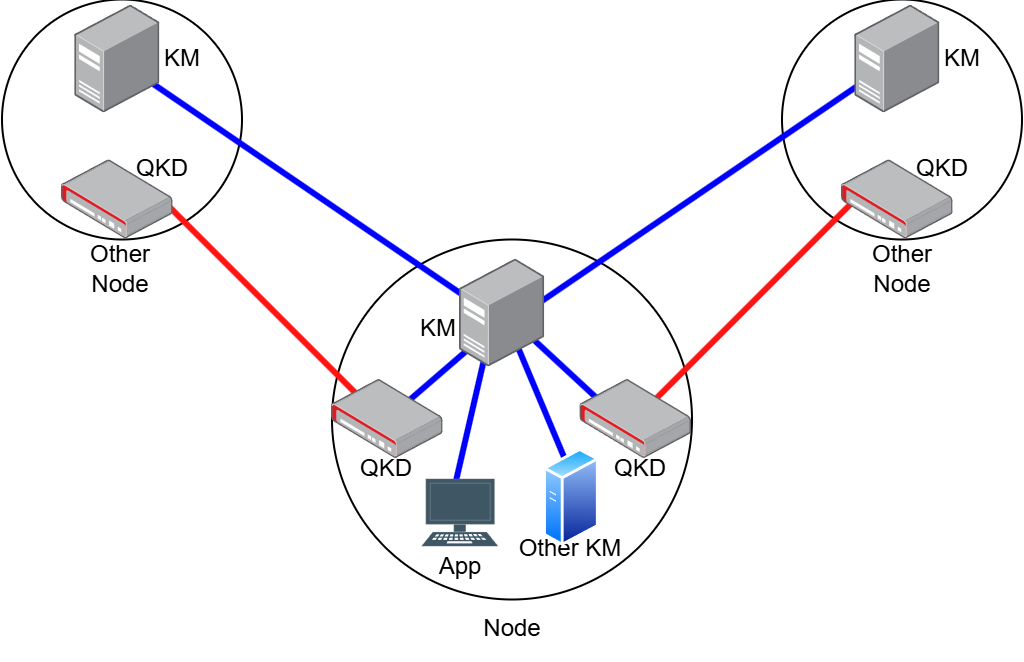}
\caption{Node model. A node is abstracted as a secure location consisting of one or more QKD links to other nodes and a Key Manager. The Key Manager may be connected to another domain's KM. A Key Manager may be connected to Applications that can request QKD keys.}
\label{fig:metadatapolicy_nodemodel}
\end{figure}

We model a metadata-aware key manager as described in Figure \ref{fig:metadatapolicy_nodemodel}. Within a node in a QKD network, we consider a Key Manager (KM) software which, for completeness, fulfills the roles of Key Management Agent (KMA), Key Supply Agent (KSA), and inter-domain connector (whether these are separate components/services or not, is beyond the scope of this paper; each role is optional, but the KM should fulfill at least one of the roles). As part of its KMA role, the KM receives, stores, and performs operations on keys generated via QKD links, leveraging interfaces with the corresponding KMs from the connected nodes. As part of its KSA role, the KM supplies keys to applications via an ETSI GS QKD 014 interface. As part of its inter-domain connector, the KM interacts and exchanges keys and commands with another domain's KM within the same node, via an ETSI GS QKD 020 interface. In a SDN-based network, the KM may also interface with a SDN Controller / KMS Orchestrator service (which may provide, for example, key forwarding commands). It is of note that in practice, the QKD and the classical KM may optionally be implemented as a single device.

Within the KM, any key operation is treated as an independent event with specific properties. Operations may include: key generation, key transfer/forwarding via trusted relay, key supply to an application, key voiding, encryption or decryption of a key, key supply to a cross-domain KM, key binding to a new owner, k-of-n key path combination, key splitting, etc. Each event includes references to related events. For example, in the case of a key obtained by combining or XORing two keys, references to the two events in which the two keys were generated/obtained would be used; the two events may further reference other events, so events in general will form a directed reference graph. Each event further includes time information, a digest over a canonical event representation, and a digital signature of the node over the event. All events that are created in a node are stored in a local per-node database. The exact structure of this database (engine, backups, size limit, ring buffer, etc) is outside the scope of this paper. We will assume events are stored for a minimum time for auditing purposes.

Node signatures are based on a hierarchical Certificate Authority structure: node certificates are issued/signed by domain-specific authorities; domain-specific authorities are issued/signed by international authorities, etc.

\subsection{Metadata transport and lifecycle}

While event information should be preserved in full for audit purposes, different actors require different levels of verbosity when metadata is to be shared. Moreover, it is expected that nodes and domains may have standing policies as to which information to share about the topology and working of the network. Thus, metadata events shall be exported as views. A metadata view may range from the full event information, to a signed encoded summary of the event in the interest of minimizing metadata size, and to a zero-knowledge proof on individual or combined properties of events or nodes involved. In the remainder of this article, when discussing metadata in transit, we will be referring to such audience-adjusted views.

As a base transport layer, each node shall expose a metadata access API to its interfaces (applications, other KMs in the same domain, SDN controller, cross-domain KM). Through this API, events and key metadata can be interrogated. It is expected that in some cases, an actor requiring a metadata event whose authoritative owner is some KM, cannot directly access the KM's API, posing a direct threat to the network scalability (for example, consider the case where an application connected to a QKD node in Domain A, requires security information about a key exchange that occurred in a QKD node in Domain B; even though all of Domain B nodes may share a common (virtual) local area network, and all of Domain A nodes may do the same, there is not necessarily an Internet connection between a non-border Domain A node and a non-border Domain B node). To mitigate this, we envision two possible solutions: 1) a public, static-IP per-domain authoritative server is set up to aggregate metadata from all domain nodes and to expose views over it via a public API. Several issues arise with this solution: it would require global orchestration to organize and distribute each domain's QKD authoritative metadata servers; more importantly, a domain's authoritative metadata server would become a single point of failure and a central attack vector. 2) a distributed solution in which node KMs are able to forward (views of) metadata to one another and across domains. We consider this the most scalable solution, since not only it does not present a single point of failure in any domain, but additionally, as long as a node or KM is able to forward metadata to other KMs, it is not required to also understand it, making it practical and version agile even with partial support.

The distributed option benefits from partial support of existing key supply standards. A requesting application may include an \code{Python}{extension_mandatory} or an \code{Python}{extension_optional} field (as per ETSI GS QKD 014) to request specific metadata views; the KM may include resulting metadata views or a signed bundle with a reference to the node authoritative API in the \code{Python}{key_container_extension} field of the key container response as per the same standard. Cross-domain, inter-KM communication can request and transfer metadata via the \code{Python}{extension} field, already supported by ETSI GS QKD 020. Within a domain, vendor-specific KMS or higher-level custom software could ensure the transport of metadata.

\subsection{An end-to-end example}

In our story with Alice and Bob, suppose each key operation produces an event of the following form:

\begin{lstlisting}[language=yaml]
Event:
  # Version of the common event envelope
  profile: urn:qci:event:1.0
  # Unique immutable event identifier
  event_id: urn:uuid:7c934c64-...-7adf92ebd065
  # Versioned event type
  event_type: urn:qci:type:key-transfer:1
  # Issuer object
  issuer:
    # Domain identifier
    domain_id: urn:qci:domain:D
    # Node identifier
    node_id: urn:qci:node:D:2
    # Reference or digest of the node cert
    credential_ref: sha256:123...
  # Time of occurance
  time:
    occurred_at: 2027-03-18T14:32:07.518Z
    recorded_at: 2027-03-18T14:32:07.877Z
  # Optional identifier for connecting
  # events belonging to one operation
  transaction_id: urn:uuid:1234...
  # Monotonic ordering within issuer's
  # append-only event database
  issuer_log:
    sequence: 583440
    previous_event_digest: sha256:456...
  # Keys involved in this event
  key_relations:
    - key:
        namespace: urn:qci:domain:D
        key_id: 123-456-789
        event_id: urn:uuid:500...
        length_bits: 256
      role: input
    - key:
        namespace: urn:qci:domain:D
        key_id: 555-666-777
        event_id: urn:uuid:501...
        length_bits: 256
      role: input
    - key:
        namespace: urn:qci:domain:D
        key_id: 333-222-111
        event_id: urn:uuid:7c93...
        length_bits: 256
      role: output
  # Event specific structure
  properties:
    operation: xor
    next_node: urn:qci:node:D:4
  # How to handle this event
  handling:
    must_understand: false
    cleartext_scope: domain:D
    on_scope_exit: forward-full
    evidence_scope: end-user
    unkown_type_action: drop
  # Possible extensions
  extensions:
    mandatory: {}
    optional: {}
  event_digest: sha256:9876...
  signatures:
    - algorithm: ML-DSA-65
      signer_credential_ref: sha256:888...
      value: base64:abcd...
\end{lstlisting}

In this structure, the event structure is given by the \code{Python}{profile} version and the specific \code{Python}{event_type}. Each event is uniquely identified with an \code{Python}{event_id}. The issuer of the event identifies itself by its \code{Python}{domain_id}, \code{Python}{node_id} within the domain, and a reference (or a hard copy) of its certificate via \code{Python}{credential_ref}. The time at which the event occured and the time at which it was recorded in the logs are also included. When multiple events get generated as part of the same operation, a \code{Python}{transaction_id} identifier is included. To preserve the local ordering and ensure integrity of events, each event is attached a node-specific \code{Python}{sequence} number and includes a reference to the digest of the last event in the node database via \code{Python}{previous_event_digest}. All keys involved are described in the \code{Python}{key_relations} object; in the case of the presented event, two 256-bit input keys were XORed (as per the \code{Python}{operation} property) into one 256-bit output key, which was forwarded to \code{Python}{next_node}. Several handling criteria are included: a further node which receives this metadata is not required to understand it, as per \code{Python}{must_understand}; the metadata is to be preserved in cleartext only within the scope of the \code{Python}{domain:D} (other possible scopes could be: local-node, same-domain, named-entities, end-user). Upon scope exit, the data is to be forwarded in full (other mechanisms could be: drop, encrypt, etc.). The evidence that the event took place, even without complete data, is to be provided up to the end-user. If a node does not understand the event profile or the specific type, it is to take the action of dropping the metadata altogether (other possible actions could be: preserve, digest-only, abort). Finally, optional extensions may be supported, an event digest is added, and the node's signature is appended.

When Alice requests a key with Bob, she includes in her \code{Python}{enc_keys} ETSI GS QKD 014 request, the following mandatory extension:

\begin{lstlisting}[upquote=false]
"extension_mandatory": [
    {
        "key_metadata": {
            "version": "urn:qci:metadata:1.0",
            "transport_policy": "forward",
            "view_policy": "full"
        } 
    }
]
\end{lstlisting}

The metadata request is carried over (by virtue of the \code{Python}{forward} policy) by the KMs, within the domain via vendor proprietary KM API, and cross-domain via the \code{Python}{extension} property defined in ETSI GS QKD 020. Upon key supply, the full event chain is provided back to Alice (Alice could have selected a more minimal view policy than "full", for example by specifying only the fields she is interested in). After the security incident disclosure and inspecting the event chain, Alice now knows the exact paths which the keys have taken, the encryption methods, and potentially other useful information about nodes involved.

In this example, algorithms (SHA256, ML-DSA-65) are illustrative only. In a fully described standard, the Event data structure should be extended to cover possible event types, all event-specific properties, as well as optional per-node, per-kms, or per-domain data such as firmware version, device configuration, location, certification level etc. Additionally, extension-based requests (through ETSI GS QKD 014 / ETSI GS QKD 020) should include not only requirements on metadata sharing, but optionally constraints on specific metadata fields such as: do not use nodes in jurisdiction X; do not use nodes without a minimum certification level Y; do not use forwarding mechanisms other than XOR or non-ITS key derivation. The returned (or asynchronously-fetched) metadata would then act as evidence that the requirements were satisfied.

With this approach, the architecture provides mechanisms addressing all requirements: 1) key provenance is verifiable cryptographically by cross-checking related events; 2) key composition is transparent via event types corresponding to operations; 3) key lifecycle is traceable via the key reference graph; 4) jurisdictions involved can be inferred from node paths and domains; 5) all claims are signed in a hierarchical manner, making verification feasible; 6) elements can be selectively disclosed by replacing individual events with zero-knowledge proof-based views; 7) in case of security incidents, impact and damage can be assessed; 8) key objects can be augmented with rights, custody, and delegation data; 9) the metadata system can be implemented at a higher-level software above the device vendors, ensuring vendor independence, while metadata can be transported and encrypted in various forms; 10) metadata grows only linearly with event actions, not exponentially; 11) algorithm and properties are versioned and can be updated as networks evolve.

\vfill

\end{document}